\documentclass[prd,10pt,twocolumn,tightenlines,preprintnumbers,showpacs,superscriptaddress,notitlepage,nofootinbib,eqsecnum,floatfix,longbibliography]{revtex4-1}
\pdfoutput=1
\usepackage[utf8]{inputenc}
\usepackage[T1]{fontenc}
\usepackage{mathrsfs}
\usepackage{bbold}
\usepackage{amsmath}
\usepackage{amssymb}
\usepackage{mathtools}
\usepackage{amsfonts,dsfont}
\usepackage{array}
\usepackage{bm,bbm}
\usepackage{graphicx}
\usepackage{xcolor}
\usepackage{enumitem}
\usepackage{soul}
\usepackage{stmaryrd}
\usepackage{hyperref}
\usepackage{cancel}
\usepackage{tikzsymbols}
\usepackage{diagbox}
\usepackage[normalem]{ulem}
\usepackage{lipsum, babel}
\usepackage{algorithm}
\usepackage{algpseudocode}

\usepackage{comment}

\hypersetup{
    colorlinks=true,     
    linkcolor=blue,      
    citecolor=blue,      
    filecolor=blue,      
    urlcolor=blue        
}

\newcommand{\mc}[1]{\mathcal{#1}}

\newcommand{\id}{\mathbb{1}}

\newcommand{\sx}{\sigma^x}
\newcommand{\sy}{\sigma^y}
\newcommand{\sz}{\sigma^z}
\newcommand{\eqnref}[1]{Eq.~(\ref{#1})}

\newcommand*{\ket}[1]{\left|{#1}\right\rangle}

\newcommand*{\braopket}[3]{\left\langle {#1} \left| {#2} \right| {#3} \right\rangle}

\makeatletter
\def\l@subsubsection#1#2{}
\makeatother

\newcommand{\ithems}{RIKEN iTHEMS,
	Wako, Saitama 351-0198, Japan}
\newcommand{\lbnl}{Nuclear Science Division,
	Lawrence Berkeley National Laboratory,
	Berkeley, California 94720, USA}
\newcommand{\linkedin}{LinkedIn Corporation,
	Sunnyvale, California 94085, USA}
\newcommand{\ucb}{Department of Physics,
	University of California,
	Berkeley, California 94720, USA}

\begin{document}
\title{Gate Based Implementation of the Laplacian with BRGC Code for Universal Quantum Computers}

\author{Ermal Rrapaj}
\affiliation{\ucb}
\affiliation{\lbnl}
\email{ermalrrapaj@berkeley.edu}

\author{Kenneth S. McElvain}
\affiliation{\ucb}
\affiliation{\lbnl}

\author{Chia~Cheng~Chang}
\affiliation{\ucb}
\affiliation{\lbnl}
\affiliation{\linkedin}

\author{Yantao Wu}
\affiliation{\ithems}
\affiliation{\ucb}

\author{Andr\'e Walker-Loud}
\affiliation{\lbnl}
\affiliation{\ucb}

\newcommand{\alert}[1]{\textbf{\color{red}{#1}}}
\renewcommand{\vec}[1]{\boldsymbol{#1}}
\newcommand{\norm}[1]{\left\lVert#1\right\rVert}

\begin{abstract}
We study the gate-based implementation of the binary reflected  Gray code (BRGC) and binary code of the unitary time evolution operator due to the Laplacian discretized on a lattice with periodic boundary conditions. We find that the resulting Trotter error is independent of system size for a fixed lattice spacing through the Baker-Campbell-Hausdorff formula. We then present our algorithm for building the BRGC quantum circuit. For an adiabatic evolution time $t$ with this circuit, and spectral norm error $\epsilon$, we find the circuit cost (number of gates) and depth required are $\mc{O}(t^2 n A D /\epsilon)$ with $n-3$ auxiliary qubits for a system with $2^n$ lattice points per dimension $D$ and particle number $A$; an improvement over binary position encoding which requires an exponential number of $n$-local operators.  
Further, under the reasonable assumption that $[T,V]$ bounds $\Delta t$, with $T$ the kinetic energy and $V$ a non-trivial potential, the cost of  QFT (Quantum Fourier Transform ) implementation of the Laplacian scales as $\mc{O}\left(n^2\right)$  with depth $\mc{O}\left(n\right)$ while BRGC scales as $\mc{O}\left(n\right)$, giving an advantage to the BRGC implementation.
\end{abstract}

\preprint{N3AS-22-020}

\maketitle

\section{INTRODUCTION}
\label{sec:introduction}

Non-relativistic quantum systems satisfy the Schr\"odinger equation.
A few simple systems can be solved analytically, but most interesting systems, including many-body systems  common to various physics domains, require numerical solutions.
The simulation of quantum systems through quantum computers, has been proposed since the early 1980's~\cite{Manin:1980,Feynman:1982}.
Since then,
a variety of quantum algorithms for quantum simulation have been developed~\cite{Aharanov:2003,Berry:2007,Childs:2010,Wiebe:2010,Childs:2010_2,Poulin:2011,Wiebe:2011,Berry:2012}, with applications ranging from quantum chemistry to quantum field theories to spin models~\cite{Lidar:1999,Guzik:2005,Jordan:2012,Heras:2014}, many of which  make use of the Lie-Trotter-Suzuki product formula~\cite{Suzuki:1991}.

A typical choice for numerical solutions is to discretize a continuous system onto a uniform lattice, representing the Hamiltonian in the corresponding discrete basis.  The resulting discretization error can be managed by a sufficiently small lattice pitch~\cite{Kivlichan:2017}, a higher-order finite difference operator~\cite{Li:2005,Kivlichan:2017},  or via continuum extrapolation, keeping the error smaller than other sources of error. 
Despite using arbitrarily high finite difference
formulas, there are situations where convergence to the continuum is
guaranteed only for lattice spacings exponentially small in the target
error~\cite{Li:2005}. 
In this work we focus on scenarios where this does not occur. For instance, for effective field theories (EFT) on the lattice one can impose momentum cutoffs in order to avoid pathological situations. 

In quantum computing applications, the implementation of these more sophisticated operators requires the use of additional gates, which can introduce quantum-hardware sources of error.
Given the complexity and many challenges of building quantum computers, it is therefore paramount to design efficient implementations of the Schr\"odinger equation to make the best use of near-term and future hardware.

The Binary Reflected Gray Code (BRGC) encoding can provide significant simplification for quantum computing as it yields simpler spin Hamlitonians, to be simulated by qubits.
For instance, it has been proposed for encoding ladder states in $d$-level systems in gate based quantum computing \cite{Sawaya2020,DiMatteo:2020dhe}, and to map the Schr\"odinger equation in position space, in any dimension, to the $XZ$ model~\cite{Chang:2022}.

In this article, we analyze the space and time complexities associated with the time evolution operator in BRGC as well as in the binary position encoding, useful both in studies of quantum dynamics and ground state properties, via adiabatic ground state algorithms, of a many-body quantum system.
For this comparison the exponential of the kinetic energy operator, or equivalently, the Laplacian, is represented as a composition of single qubit rotations, controlled rotations, and Toffoli (CCNOT) gates.
The Laplacian is then rendered as a sum of products of simple operators in both encodings (Sec.~\ref{sec:Laplacian}).
We compute the Trotter error for both encodings (Sec.~\ref{sec:trotter_error}) and provide the algorithm for the BRGC Laplacian (Sec.~\ref{sec:gates}).

The resulting circuit is seen to scale linearly with the number of qubits for BRGC and exponentially for binary code. As a numerical illustration, we display the results of performing adiabatic evolution with gate-based universal quantum computers in Sec.~\ref{sec:hardware} and compare with the quantum fourier transform (QFT) implementation~\cite{Nielsen:2000,CooperSmith:2002}.
A summary of our findings and conclusions may be found in Sec.~\ref{sec:summary}.

\section{Laplacian}
\label{sec:Laplacian}

This work employs the centered first order finite difference representation of the Laplacian.%
\footnote{It is worth exploring the efficacy of BRGC with higher order finite difference expressions~\cite{Li:2005} and 3D stencils for the Laplacian operator~\cite{o2006family}, but we leave that to future work.} %
Periodic boundary conditions are assumed with the goal of simplifying the construction of the approximate Hamiltonian.  Non-periodic boundary conditions can be simulated by choosing the lattice volume to be appropriately larger than the dynamically accessible space for the computation.

We begin by summarizing the representation of the Laplacian matrix in both BRGC and binary encodings.
The equation is discretized on a lattice with spacing (pitch) $a$ and $N$ positions in each of the $D$ directions.
Then, up to a discretization error proportional to $a^2$, the Laplacian becomes an $N \times N$ matrix which acts on the
wavefunction at the discrete points $\mathbf{x}=a\mathbf{r}, \mathbf{r}\in \mathbb{Z}^D$
\begin{equation}
  \begin{split}
  \nabla^2 \psi(\mathbf{x}) &\approx  \frac{1}{a^2}(L\psi)(a\mathbf{r}) \\
  &= \frac{1}{a^2}\left[ \left( \sum\limits_{\mathbf{m} \in \mc{N}(r)} \psi(a\mathbf{m})\right) - 2D \psi(a\mathbf{r})\right] ,
\end{split}
\end{equation}
with $\mc{N}(\mathbf{r})$ indicating the set of nearest neighbors around the lattice site $a \mathbf{r}$.
Here,  $L$ is used to denote the dimensionless Laplacian.
For simplicity of notation, throughout this article, we will work in natural units, $\hbar=c=1$.
Higher order finite difference formulas could also be implemented (see~\cite{Kivlichan:2017} for an in-depth analysis), at the cost of higher connectivity in the resulting Laplacian matrix.
The discrete Schr\"odinger equation then follows
\begin{equation}
	(H\psi) (a\mathbf{r}) = -\frac{(L\psi)(a\mathbf{r})}{2 M a^2}   + (V\psi)(a\mathbf{r}) = E \psi(a\mathbf{r}).
	\label{eq:Hamiltonian}
\end{equation}
The next step is to encode the positions in states of qubits. Following our previous work~\cite{Chang:2022}, positions are associated with $n$-body qubit states in the computational basis, and we identify each basis state's amplitude with the wave function at the corresponding point. Alternatively, one could associate position $i$ with a set of qubits~\cite{Abel2021,Pilon2021}, but the number of qubits, in that case, would be comparable to the number of classical bits (no storage advantage). The Laplacian for $A$ particles in $D$ dimensions can be constructed from the single particle Laplacian in one dimension through Kronecker sum, and in the rest of this work we focus on this case and generalize to multiple particles in our final results.

\subsection{Notation}
\label{subsec:notation}
We define qubit (spin) projection operators
\begin{equation}
  P^0 = \frac{\id + \sz}{2} = \begin{bmatrix}1&0\\0&0\end{bmatrix},\quad  P^1 = \frac{\id - \sz}{2} = \begin{bmatrix}0&0\\0&1\end{bmatrix}
\end{equation}
where $P^0$ projects onto $\ket{0}$ (spin up)  and $P^1$ onto $\ket{1}$ (spin down) for a single qubit.
Raising and lowering operators on a spin are defined as
\begin{align}
  \sigma^+ =& \frac{\sigma^x + i \sigma^y}{2} = \sx P^1 = \begin{bmatrix}0&1\\0&0\end{bmatrix},\nonumber \\
  \sigma^- =& \frac{\sigma^x - i \sigma^y}{2} = \sx P^0 = \begin{bmatrix}0&0\\1&0\end{bmatrix}.
\end{align}

The variable $n$ will indicate the number of qubits in the system and $O_i$ indicates that the operator $O$ is acting on qubit at location $i$. When the value of $n$ is clear from the context, it will be omitted. For example,
\begin{equation}
    n=4 \longrightarrow \sz_1 = \id \otimes \id \otimes \sz \otimes \id
\end{equation}

\subsection{Binary encoding}
\label{subsec:binary-code}
The simplest form of the discrete Laplacian is given by the nearest-neighbor finite difference method,
\begin{equation}
    \frac{\partial^2}{\partial x^2}f(x) = \frac{1}{a^2}\left[f(x+a) + f(x-a) - 2f(x)\right]
\end{equation}
The binary representation of the Laplacian, $L_{0...n-1}^{(n,\text{bin})}$, with periodic boundary conditions can be rewritten as:
\begin{equation}
  \begin{split}
  L_{0...n-1}^{(n, \textrm{bin})}
  =& \sum_{l=1}^{n}\mc{B}_l\\
  \mc{B}_l=&(\sigma_{l{-}1}^x {-}1)  (\prod\limits_{i=0}^{l{-}2} \sigma_i^x  P_i^0 + \prod\limits_{i=0}^{l{-}2} \sigma_i^x  P_i^1)\\
  \mc{B}_{1}=&L^{(1)}_{0} = 2\sx_0
  \end{split}
  \label{eq:L_recursion}
\end{equation}
Unfortunately, the operators in the product $\prod\limits_{i=0}^{l{-}2} \sigma_i^x  P_i^0$ can not be directly implemented in gate based quantum hardware.
The product is decomposed into a sum of direct products of Pauli matrices. For instance,
\begin{align}
\begin{split}
L_{0,1}^{(2,\textrm{bin})}&=
    \sigma_1^x+\sigma_0^x\sigma_1^x,
\\
L_{0...2}^{(3,\textrm{bin})}&=
    \sigma_2^x + \frac{1}{2}L_{0,1}^{(2,\textrm{bin})} \sigma_2^x
    +\frac{1}{2}\sigma_1^y\sigma_2^y
    -\frac{1}{2}\sigma_0^x\sigma_1^y\sigma_2^y.
\\
\end{split}
\end{align}
The number of $\rm n-$local operators generated by this procedure is  $2^{(n-2)}$ for $n\geq 2$. The large number of multi qubit operators will inevitably generate rather deep circuits when decomposed into single and two qubit operations. If the product of the projection operators can be applied without being first decomposed into Pauli strings (product of Pauli matrices), the resulting circuit depth would be much lower. This is the case for BRGC, which we discuss in the next sections.

\subsection{BRGC encoding}
\label{subsec:brgc_code}
Any Gray encoding of the positions guarantees that neighboring bit-strings differ in exactly one bit.
Gray code is an alternative compact binary encoding of integers $0$ to $2^n{-}1$ into $n$ bits.
The recursion equation for BRGC encoded Laplacian from~\cite{Chang:2022} can be rewritten as an iterative sum,
\begin{equation} \label{eqn:graylap}
 L^{(n,\textrm{BRGC})}_{0\ldots n-1}=\sum_{k=0}^{n-1} \mc{G}_k\quad\quad\quad\quad
\end{equation}
\begin{equation} \label{eq:G_k}
\begin{split}
    \quad\mc{G}_k &= \left(\sigma_{k}^x - \sigma_{k-1}^x\right) \prod\limits_{i = 0}^{k-2} {P_i^0}\\
  \quad\mc{G}_{0} &= L^{(1)}_{0} = 2\sx_0
\end{split}
\end{equation}
Then, for $n=2$, the neighbor contributions to the Laplacian operator are
\begin{equation}
\label{eq:gray2lapbase}
\begin{split}
    L^{(2,\textrm{BRGC})}_{0, 1} &= \mc{G}_1 + 2 \sigma^x_0 \\
    &= \sigma^x_1 + \sigma^x_0
\end{split}
\end{equation}
which is recognized as the transverse Hamiltonian acting on qubits $0$ and $1$.
As is shown in the following sections, $\exp(i \lambda \mc{G}_k)$, with $\lambda=\Delta t / (2 M a^2)$, can be expressed exactly on hardware, and no further Trotter expansion is required.

\section{Trotter approximation error}
\label{sec:trotter_error}
\begin{figure}[t]
\includegraphics[width=\columnwidth]{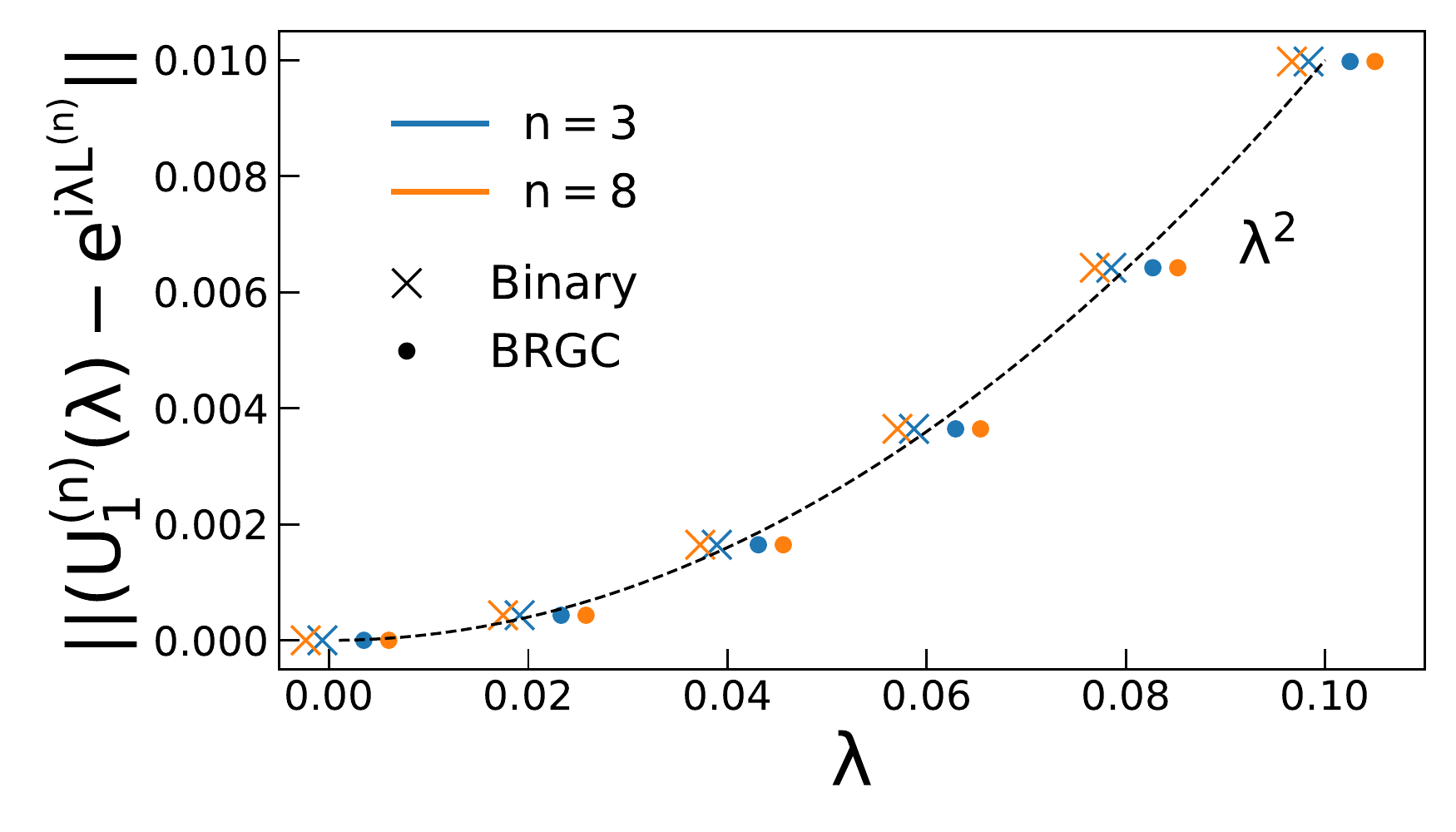}
\caption{The spectral norm of the difference between the $\rm 1^{st}$ Trotter approximation and the exact value as a function of $\lambda$ is shown for system sizes $3 \leq n \leq 8$. The dashed line serves as guide for understanding the quadratic error scaling as function of the step size $\lambda$. There is no dependence of the approximation error on system size $n$. The various markers have been slightly shifted for visibility. For the binary code, each individual Pauli string has been exponentiated separately in $U_1$ (see discussion in Sec.~\ref{subsec:binary-code})}
\label{fig:fidelity}
\end{figure}

The time evolution can be expressed as an accumulation of tiny time steps,
and the propagator can be approximated by the Trotter-Suzuki expansion. To first order, the expression is,
\begin{equation}
 \begin{split}
  e^{-i\Delta t H} &=
    e^{-i\Delta t V}\ e^{i \frac{\Delta t   L}{2 M a^2}}
    + \mc{O}\left((\Delta t)^2\right)
    \\
 \end{split}
\end{equation}
Higher order formulas are also available and are left for future study~\cite{Suzuki:1993,Hatano:2005,Wiebe:2010}.
Here we focus on the one dimensional, single particle BRGC Laplacian in $n$ qubit representation,
\begin{equation}
 \begin{split}
  \exp \left( i \lambda   L^{(n)} \right)
  &=
  \prod_{k=0}^{n-1} \exp \left(i\lambda \mc{G}_k\right)
    +\mc{O}\left(\lambda^2\right) \\
  &\equiv
  U^{(n)}_1(\lambda)+ \mc{O}\left(\lambda^2\right)
 \end{split}
\end{equation}
An upper bound on the spectral norm error, based on Ref.~\cite{Childs:2021}, is provided,
\begin{equation}
 \begin{split}
  \norm{U^{(n)}_1(\lambda) - e^{i \lambda   L_{0...n-1}^{(n)}}} \leq& \frac{\lambda^2}{2} \sum^{n-1}_{j=2} \norm{ \sum^{j-1}_{k=0} \left[ \mc{G}_j,\mc{G}_k\right] }\\
  =& \frac{\lambda^2}{2} \sum^{n-1}_{j=2} \norm{
    \left[ \mc{G}_j,L^{(n)}_{0...j-1}\right] }
\end{split} .
\label{eq:Trotter_error_prediction}
\end{equation}
To evaluate this expression, we begin with the commutators, $\left[ \mc{G}_j, \mc{G}_k \right]$ and expand with \eqnref{eq:G_k},
\begin{align*}
\left[ \mc{G}_j, \mc{G}_k \right] &=
    \left[ (\sx_j - \sx_{j-1}) \prod_{i=0}^{j-2} P_i^0,
        (\sx_k - \sx_{k-1}) \prod_{i^\prime=0}^{k-2} P_{i^\prime}^0
    \right]\, .
\end{align*}
Note that operators on different qubits commute, and the projectors satisfy the relation $(P_i^{0,1})^2 = P_i^{0,1}$. The non-vanishing contributions from the commutator occur between the projection operators and  $\sigma^x$,
\begin{equation*}
\left[ P_i^0, \sx_k \right] = -i \delta_{i,k} \sy_k\, .
\end{equation*}
Two cases are needed to evaluate $\left[\mc{G}_j, \mc{G}_k\right]$,
$k = j-1$ and $k \le j-2$.  For $k=j-1$,
\begin{gather} \label{eqn:keqjm1}
\left[
   \left(\sigma^x_j - \sigma^x_{j-1}\right)\prod\limits_{i=0}^{j-2} P_i^0 ,
  \left(\sigma^x_{j-1} - \sigma^x_{j-2}\right)\prod\limits_{i=0}^{j-3} P_i^0
\right] \\
= i\left(\sigma^x_j - \sigma^x_{j-1} \right) \left(\prod\limits_{\begin{matrix} i=0 \\ i\ne j-2\end{matrix}}^{j-2} P_i^0 \right) \sy_0, \nonumber
\end{gather}
where single operators on distinct bits have been factored out.
Note that in the first line, the term $\sigma^x_{j-1}$ in the right of the commutator has an index greater than all those in the projector on the left, and therefore makes no contribution.
The full projector product with the interacting qubit index explicitly canceled out is retained for uniformity with the remaining $k \le j-2$ case which is slightly more complicated:
\begin{gather} \label{eqn:eqnklejm2}
\left[
  \left(\sigma^x_j {-} \sigma^x_{j{-}1} \right) \prod\limits_{i=0}^{j{-}2} P_i^0 ,
  \left(\sigma^x_{k} {-} \sigma^x_{k{-}1}\right)\prod\limits_{i=0}^{k{-}2} P_i^0
\right] = -i \left(\sigma^x_j {-} \sigma^x_{j-1}\right) \\
 \times \left( \left(\prod\limits_{\begin{matrix} i=0 \\ i\ne k \end{matrix}}^{j-2} \!\! P_i^0\right)  \sy_k  -
\!\! \left(\prod\limits_{\begin{matrix} i=0 \\ i\ne k-1\end{matrix}}^{j-2} \!\! \!\!\! P_i^0\right) \sy_{k-1} \right). \nonumber
\end{gather}
In a sum of these commutators for fixed $j$, \eqnref{eqn:keqjm1} can be seen to cancel the left hand side of the result for \eqnref{eqn:eqnklejm2} with $k=j-2$.
This pattern repeats itself with the right hand side of \eqnref{eqn:eqnklejm2} for $k$ canceling the left hand side for $k-1$,
leaving only the right hand side of the $k=0$ case uncanceled.   The sum for fixed $j$ is then
\begin{align}
\sum_{k=0}^{j-1} \left[ \mc{G}_j, \mc{G}_k \right] &=
    i(\sx_j- \sx_{j-1}) \left(\prod_{i=1}^{j-2} P_i^0 \right)\sy_{0} ,
\end{align}
which we also confirmed through explicit evaluation. This expression has a spectral norm of 2.
Together with Eq.~(\ref{eq:Trotter_error_prediction}) we can determine an upper bound on the error,
\begin{equation}
\begin{split}
  \norm{U^{(n)}_1(\lambda) - e^{i \lambda   L_{0...n-1}^{(n)}}}
  &\leq \frac{\lambda^2}{2} \sum^{n-1}_{j=2} \norm{ \sum^{j-1}_{k=0}
    \left[ \mc{G}_j,\mc{G}_k\right] }\\
  &= \frac{\lambda^2}{2} \sum^{n-1}_{j=2} 2\\
  &= (n-2)  \lambda^2.
\end{split}
\label{eq:Trotter_error_v1}
\end{equation}
This expression suggests that the Trotter error scales linearly $n$ and quadratically with $\lambda$. However, this bound is very conservative, and as we show, the error does not depend on system size at second order in $\lambda$.
\begin{gather}
  U^{(n)}_1(\lambda) - e^{i \lambda   L_{0...n-1}^{(n)}} = \frac{\lambda^2}{2} \sum^{n-1}_{j=2}  \sum^{j-1}_{k=0} \left[ \mc{G}_j,\mc{G}_k\right] +\mathcal{O}(\lambda^3) \nonumber \\
  = i\frac{\lambda^2}{2} \sum^{n-1}_{j=2}   \left(\sigma_{j}^x {-} \sigma_{j-1}^x\right)  \prod\limits_{i=1}^{j-2}{P_i^0} \sigma_0^y
  +\mathcal{O}(\lambda^3)
\label{eq:Trotter_error_v2}
\end{gather}
Now, consider applying the operator to a basis state.   There is a last projector admitting the state, say it is for $k=K$.
Then all earlier projectors will also admit the state and for $k > K$ the projectors will not.    For that state we can reduce the
operator to
$i\left(\sigma_{K}^x - \sigma_1^x\right)\sigma_0^y$,
which has spectral norm of 2. Thus, the entire sum has norm 2
\begin{equation}
\begin{split}
\norm{\sum^{n-1}_{j=2}   \left(\sigma_{j}^x {-} \sigma_{j-1}^x\right)  \prod\limits_{i=1}^{j-2}{P_i^0} \sigma_0^y}
&= 2
\end{split}
\label{eq:Trotter_error_prediction_result}
\end{equation}
Putting this together with \eqnref{eq:Trotter_error_v2}, and using the Baker-Campbell-Hausdorff formula with \eqnref{eq:Trotter_error_prediction_result}, we find that the spectral norm of the difference between the $\rm 1^{st}$ order Trotter expansion and the exact time evolution operator is independent of $n$, in sharp contrast with the linear scaling one would expect from Eq.~(\ref{eq:Trotter_error_v1}),
\begin{align}\label{eq:Trotter_error}
\norm{U^{(n)}_1(i \lambda) {-} e^{i \lambda   L^{(n)}} }
    &=
    \frac{\lambda^2}{2} \norm{ \sum^{n-1}_{b=2}
    \left[ \mc{G}_b,\mc{L}^{(n)}_{0...(b-1)}\right]}
    + \ldots
\nonumber\\&=
    \lambda^2 + \mc{O}(\lambda^3)\, .
\end{align}
In Fig.~(\ref{fig:fidelity}), we plot the spectral norm of the difference $\rm 1^{st}$ order Trotter expansion and the exact time evolution operator, verifying \eqnref{eq:Trotter_error}.
As can be seen, in both codes, there is quadratic dependence on $\lambda$ and no dependence upon $n$.
\footnote{While an analogous version of Eq.~(\ref{eq:Trotter_error}) applies to the binary encoding, since we can not directly apply it on hardware through gates, we have expanded each term through a product of Pauli matrices, and computed the error shown in Fig.~(\ref{fig:fidelity}) by exponentiating each product separately. It is a rather remarkable result than Trotter error in this case matches the BRGC result.}

Therefore, the overall cost in terms of the number of Trotter steps is given by
\begin{equation}
 \begin{split}
  C = \mc{O}(T^2 A D /\epsilon)C_{gate},\\
 \end{split}
\end{equation}
where the effects of the particle number $A$, the dimension number $D$, and the spectral norm error $\epsilon$ are included. The next section  expresses the gate representation and analyses the depth and width of the resulting circuit.

\section{Gate Implementation}
\label{sec:gates}
\begin{figure}
 \includegraphics[width=\columnwidth]{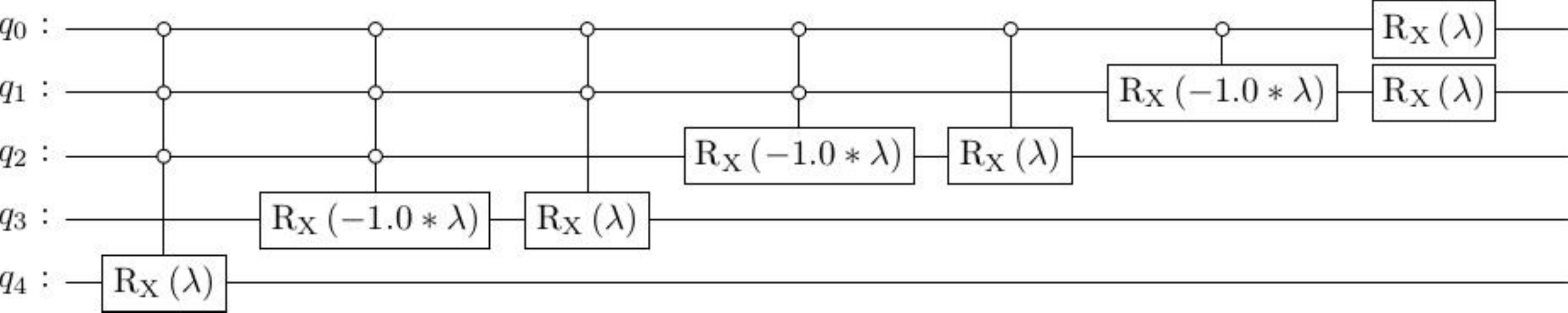}
 \caption{Circuit representation of $U^{(5)}_1(\lambda)$ with multi qubit controlled rotations, see the text for a description.}
 \label{fig:U1A_general_circuit}
\end{figure}

As the binary encoding inevitably leads to deep networks due to the exponential number of $n-$local operators, here we focus on the BRGC representation in the hope for efficient shallower circuits.
To achieve this goal, a couple of considerations about $\mc{G}_l$ will prove useful,
\begin{itemize}
 \item The Pauli operators $\sigma_{n-1}^x$ and $\sigma_{n-2}^x$ commute. In other words, they can be applied in any order without inducing additional errors.
 \item The product of projection operators, $\prod\limits_{i = 0}^{n - 3} {P_i^0}$, is nonzero (equal to $1$) only if each of the indexed qubits $i\leq n-3$ is equal to $0$.
\end{itemize}
Based on the considerations above, $\exp(i\lambda \mc{G}_l)$ is an $l-2$ controlled two qubit rotation.
Fig.~(\ref{fig:U1A_general_circuit}) plots the circuit representation of $U^{(n=5)}_1(\lambda)$. Each vertical line in the figure represents a multi-qubit controlled rotation gate, with the white circles denote the control qubits. If the NISQ hardware could implement such multi control rotation in a single step, then the circuit depth would be $2(n-1)+1=2n-1$ and the width would be $n$. Unfortunately, a single step implementation is not currently available, but it may be so for ion based quantum computers in the near future~\cite{Katz:2022}.

Given one- and two-qubit operations,
based on the multi-qubit control not gate representation in~\cite{Nielsen:2000}, then the circuit can be decomposed  as shown in Algorithm~\ref{alg:cap}.
\begin{algorithm}[H]
\caption{BRGC Laplacian circuit}
\label{alg:cap}
\begin{algorithmic}
\State $n_a=n-3,\ n_{\rm tot}=2n-3, n_c = n-2$
\State ${\rm CCX}_{00}(0,1,n_{\rm tot}-1)$
\For{$(i = 2,\ i<n_c,\ i++)$}
    \State ${\rm CCX}_{10}(i,n_{\rm tot}-i+1,n_{\rm tot}-i)$
\EndFor
\State{$c=0$}
\For{$(j=3,\ j<n,\ j++)$}
    \State{$C_{R_X,1}(\lambda,n_{\rm tot}-n_c+1-c,n-1-c)$}
    \State{$C_{R_X,1}(-\lambda,n_{\rm tot}-n_c+1+c,n-2-c)$}
    \If{$(n_{\rm tot}-n_c+2+c)\%n_{\rm tot}>n-1$}
        \State{\fontsize{8.5pt}{2} \selectfont ${\rm CCX}_{01}(n_c-1-c,(n_{\rm tot}-n_c+2+c)\%n_{\rm tot},n_{\rm tot}-n_c+1+c)$}
    \Else
        \State{\fontsize{8.5pt}{2} \selectfont ${\rm CCX}_{00}(n_c-1-c,(n_{\rm tot}-n_c+2+c)\%n_{\rm tot},n_{\rm tot}-n_c+1+c)$}
    \EndIf
    \State{$c\ +=\ 1$}
\EndFor
\State{$C_{R_X,0}(\lambda,0,2)$}
\State{$C_{R_X,0}(-\lambda,0,1)$}
\State{$R_X(\lambda,0)$}
\State{$R_X(\lambda,1)$}
\end{algorithmic}
\end{algorithm}
\noindent
${\rm CCX}_{a,b}(i,j,k)$ is a Toffoli gate which acts on qubit $k$ if $\sigma_z |i \rangle=a$ and $\sigma_z |j \rangle=b$,
and $C_{R_X,a}(\lambda,i,j)$ is a controlled rotation with angle $\lambda$ on qubit $j$ if $\sigma_z |i \rangle=a$.
The Toffoli gate can be expressed through five two-qubit gates~\cite{Barenco:1995,Sleator:1995,DiVincenzo:1998,Yu:2013}, and has also been implemented directly on quantum hardware~\cite{Monz:2009,Fedorov:2012,Luo:2015}.
The Laplacian requires the application of recursively smaller gates, based on Eq.~(\ref{eqn:graylap}). The application of the initial sequence of Toffoli gates will require $n-3$ auxiliary qubits. While the time complexity will increase, the space complexity will not, as the auxiliary qubits from the first step can be re-used for the smaller subsystems.

\begin{figure}[t]
 \includegraphics[width=\columnwidth]{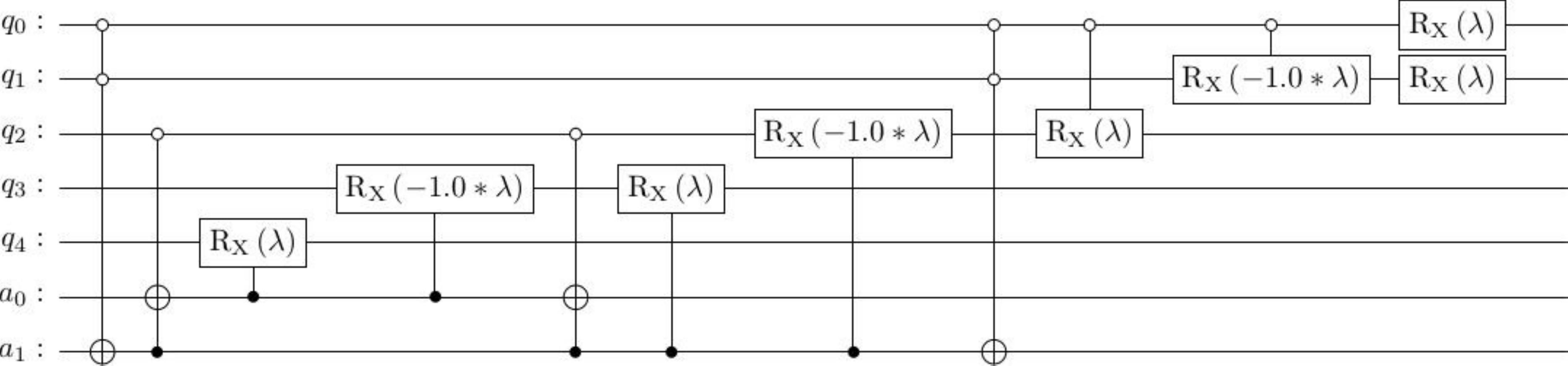}
 \caption{The implementation of $U^{(5)}_1(\lambda)$ based on sequential decomposition of the multi-qubit control NOT gate.
 \label{fig:ncrxx}}
\end{figure}

The control Toffoli gates for the lower qubits repeat during recursion and can be reordered since they commute.
The reordering exposes adjacent Toffoli gates that are identical, and for which the net effect is the identity operator.
Thus, the total circuit depth and width complexities for Algorithm~\ref{alg:cap} are
\begin{itemize}
 \item {\fontsize{9.5pt}{2}\selectfont $depth=2\left[(n-3)C_{\rm CCX}+(n-2)C_{C_{R_X}}\right]+1=\mc{O}(n)$}
 \item $\rm width=n+(n-3)=2n-3=\mc{O}(n)$
\end{itemize}
In our implementation the number of gates required is equal to the circuit depth.
Smaller systems do not require Toffoli gates. For instance, the 3 qubit system requires two controlled rotations for which the first qubit is the control, and the system of 2 qubits requires just two rotations (single qubit operations), see Eq.~(\ref{eq:gray2lapbase}).
As an illustration of the more general case, in Fig.~(\ref{fig:ncrxx}) we show the resulting circuit from Fig.~(\ref{fig:U1A_general_circuit}).
The total cost of the gate based implementation of the BRGC Laplacian is
\begin{equation}
 C^{\rm BRGC}=\mc{O}(T^2 n D A/\epsilon).
\end{equation}
It worth mentioning that this result does not account for swap gates which may be needed on hardware that can only perform local two-qubit interactions.

We can also compare with a Quantum Fourier Transform (QFT) implementation for which $\exp(i\lambda L^{(n)})$ can be implemented exactly with a respective cost of $\mc{O}(n^2 D A)$~\cite{Nielsen:2000} (it may be further reduced to $\mc{O}(n \log(n) D A)$ based on section III.B from~\cite{Su:2021}).
An approximate QFT implementation would further reduce the cost to $\mc{O}(n \log(n) D A)$~\cite{Nam:2020}. While QFT does not suffer from the approximation error associated with finite difference methods, it  generally has a higher cost in the number of qubits, and NISQ devices have a limited decoherence time~\cite{Preskill:2018}.

From the above analysis, one would conclude that the BRGC Laplacian scales better in lattice resolution, but the QFT implementation
requires fewer steps.  However, in many physical problems, the main source of error is the Trotter approximation of the time evolution operator due to the non-commutativity of the kinetic and potential terms. This error,  present in both QFT and finite difference representations, can be rather large.
As $n$ increases, the finite difference approximation error vanishes and the step size is dominated by $\left[T,V\right]$ which forces a smaller common step size regardless of the representation of the Laplacian.

As an illustration, we briefly analyze in the continuum limit the introduction of a Harmonic Oscillator (HO) potential, with
\begin{equation}
H=T+V=\frac{P^2}{2M} + \frac{M \omega^2}{2}X^2
\end{equation}
The commutator is
\begin{equation}
\begin{aligned}
\left[P^2, X^2\right]&= -4i X P - 2 \\
\end{aligned}
\end{equation}
The matrix elements in an harmonic oscillator basis are
\begin{equation} \label{eqn:PXMatrixElements}
\begin{gathered}
\braopket{i}{\left[P^2,X^2\right]}{j} = \hfill \\
\quad\quad \left(\sqrt{i(i{-}1)}\delta_{i,j{+}2} +\sqrt{j(j{-}1)} \delta_{i{+2},j}\right)\hfill
\end{gathered}
\end{equation}
where $|j\rangle$ is an eigenstate with energy $E_j=\omega (j+1/2)$.
The max eigenvalue of \eqnref{eqn:PXMatrixElements}, giving the spectral norm,  grows slightly super-linearly with the quanta cutoff $\Lambda$, with $\Lambda=10$ yielding a max eigenvalue of $\sim 11.1$, $\Lambda=100$ yielding $\sim 176.1$ and $\Lambda=1000$ yielding $\sim 1944.3$, obtained by numerical diagonalization  with cutoff $i,j\le\Lambda$ .
The discrete position space basis has a momentum cutoff of $\sim 1/a$.     In a similar way the discrete HO basis can be seen to have a momentum cutoff of
$\sim (1/b)\sqrt{\Lambda}$,  with $b = (M\omega)^{-1/2}$ the length scale of the HO~\cite{Furnstahl:2012}. 
Equating momentum cutoffs we have
\begin{equation}
1/a \approx (1/b) \sqrt{\Lambda } \Rightarrow  \Lambda \approx ( b/a)^2 = 1/(M\omega a^2)
\end{equation}

Then $\norm{\left[ P^2, X^2\right]} \approx \Lambda \approx  1/(M \omega a^2)$, and
\begin{equation}
\norm{\left[T,V\right]} = \left(\omega^2/4\right)\norm{\left[P^2,X^2\right] } \approx  \omega/(4 M a^2) ,
\end{equation}
while the error associated with the finite difference approximation of the Laplacian is $\approx 1/(2 (M a^2)^2)$ based on Eq.~(\ref{eq:Trotter_error}). It is easy to see that as oscillator frequency is increased to confine the harmonic oscillator states in the volume, the error from the commutator of the kinetic and potential terms dominates the error for $\omega > 2(Ma^2)^{-1}$, and therefore the step size, removing that advantage from the QFT implementation of the Laplacian, see the next section.
Then, the only gate cost scaling that matters in the comparison is with respect to $n$ .    The approximate QFT scaling is $\mc{O}\left(n \log(n) \right)$ and BRGC scaling is $\mc{O}\left( n\right)$, giving a clear advantage to BRGC.    As will be shown below, the BRGC circuit size, even for small $n$, is smaller, leading to lower noise circuits.

\section{Experiments on Universal Quantum Computers}
\label{sec:hardware}
As a simple illustration, we consider a particle in a box in one dimension with periodic boundary conditions with four lattice points that can be represented by $\rm n=2$ qubits. For this qubit size, there is no Trotter error in the implementation of the Laplacian for both encodings. Thus, the only source of such error is due to the non-commutativity of the potential and kinetic terms. The length of the box is chosen to be $\rm L=10\ fm$, and the potential is
\begin{equation}
 \rm V= \begin{pmatrix} V_0, & r<L/2 \\ -V_0, & r\geq L/2 \end{pmatrix},
\end{equation}
where $\rm V_0=-10\ MeV$ and the particle mass is $\rm M=140\ MeV$.
For $\rm n>2$, from Eq.~(\ref{eq:Trotter_error}) the norm the error approximation of the Laplacian is $\rm \leq 2 / (2Ma^2)^2\approx 61.9\ MeV^2$.
For the given example, the spectral norm of the commutator between the kinetic and potential terms is $\rm ||\left[T,V\right]|| = 20 MeV/(2Ma^2)\approx 111.3\ MeV^2$, independently of system size for the fixed value of the lattice spacing.
As such, it is the dominating contribution to the Trotter error for any number of qubits.
The adiabatic schedule was chosen from~\cite{Albash2018,Dong2020,Chang:2022}, and it suffices to obtain the ground state with $\rm t= 10\ MeV^{-1}$ evolution time. We opted to perform $2000$ uniform steps with $\rm \Delta t = t/2000$. To account for the discreteness of the step in gate-based computing, we opted for the Magnus expansion to first order (time integrate the time-dependent coefficients over the step size) for each step~\cite{Blanes:2009}. The initial state chosen is the transverse one, $|\Psi_0\rangle= \otimes \prod |+\rangle$. This wavefunction is an eigenstate of the kinetic term, and it reproduces the continuous adiabatic evolution setup of~\cite{Chang:2022}.
In addition, it has the added benefit of being implementable in a single step by simultaneously applying the Hadamard gate to each qubit.
For the number of steps chosen, the relative error for the expectation value of the kinetic term is less than $\rm 0.07 \%$\, and for the potential term is less than $\rm 0.016 \%$.
While the error can be further reduced by decreasing the step size, we opted to pick a number of steps that displays small error in the Trotter approximation for both encodings. In addition, for comparison, we perform the time evolution using the QFT representation.
The same number of steps were applied with comparable theoretical Trotter error.
The resulting circuits for each case are described in detail in appendix~\ref{app:circuit_compression}.

\begin{figure}[ht]
\includegraphics[width=\columnwidth]{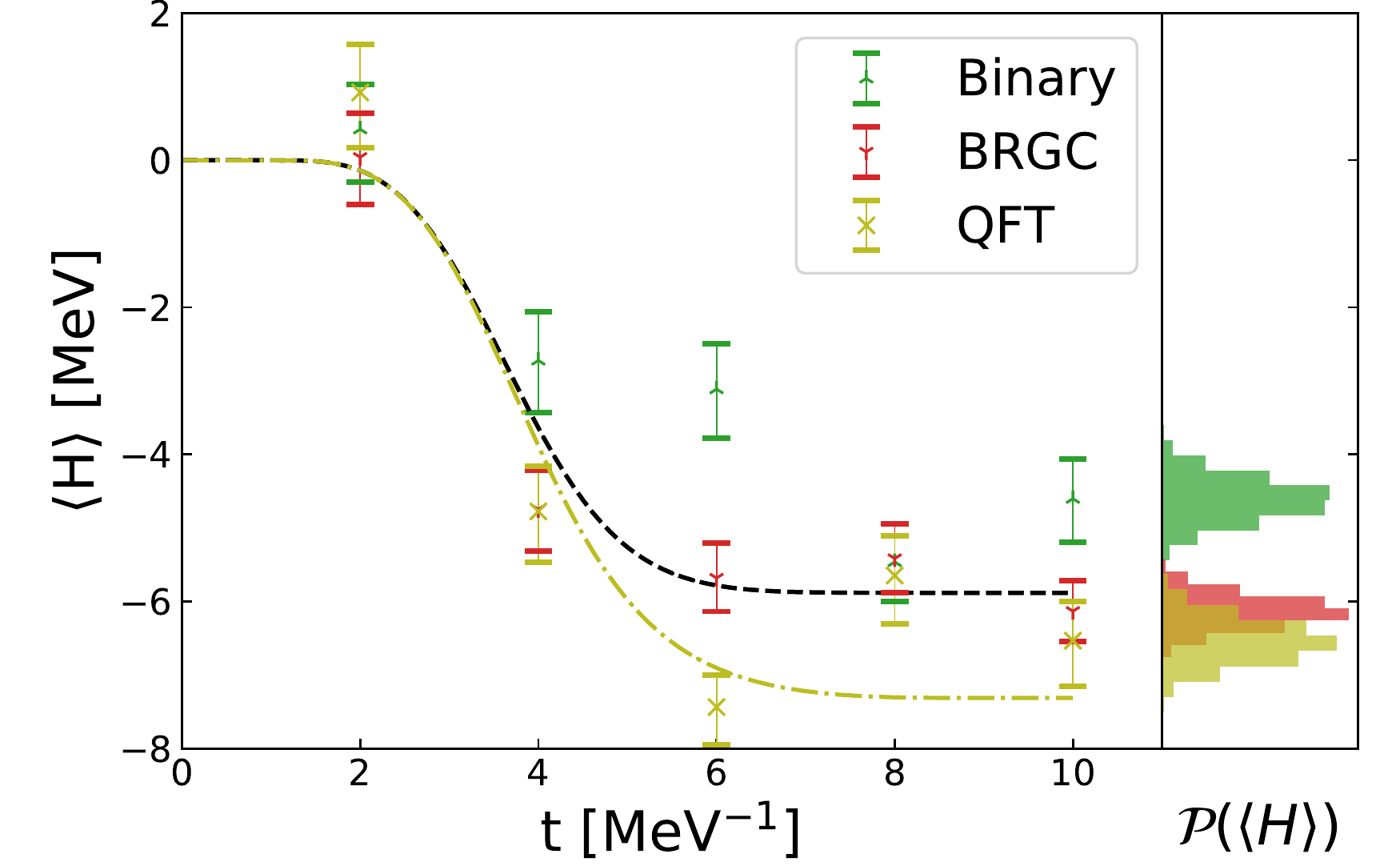}
\caption{Theoretical result from adiabatic evolution in dashed black line for the finite difference and dash-dotted yellow line for QFT. Hardware results are in red for BRGC, green for Binary encoding, and yellow for QFT. Each point is the results of 2000 samples from hardware, half for the kinetic term and half for the potential. The error bars are 95\% confidence intervals estimated with bootstrap resampling with the full bootstrap samples shown in the histograms on the right panel.}
 \label{fig:ionq_results}
\end{figure}
Fig.~(\ref{fig:ionq_results}) displays the experimental results as applied on the IonQ hardware with trapped ions~\cite{ionq}. Due to the different resolution of the Laplacian and the small number of qubits, the theoretical expectations differ between finite difference and QFT implementations. This difference can be reduced by decreasing lattice spacing, and optionally, increasing lattice size. We collected $2000$ samples for each point in the figure, half for the kinetic term and half for the potential term, and estimated the $95\%$ confidence interval (the plotted error bars) with bootstrap resampling of the samples collected. The binary encoding results do not agree with the exact values for most of the evolution schedule, including the final time point. The same happens with the QFT implementation, which also displays higher uncertainty.  The BRGC encoding, on the other hand, matches the exact result and has smaller uncertainties.

\section{Summary and Conclusion}
\label{sec:summary}
In this work we study the gate-based implementation of the BRGC and binary code Laplacian. The recursion expressions from our previous work are expressed as simple iterative sums, and the resulting Trotter error is analyzed. We find that there is no scaling of the error associated with lattice size, a rather unexpected result which we explain through the Baker-Campbell-Hausdorff formula.
We describe the algorithm for building the quantum circuit and show that both the cost and depth scale linearly with the number of qubits.

In many instances the Trotter error is dominated by the commutator $\left[T,V\right]$, thus determining the step size. In such cases, the cost of the BRGC Laplacian is better than even the approximate QFT implementation by a factor of $\log(n)$, with $n$ being the number of qubits representing the position.

As a final step, we implement our algorithm on trapped ion quantum hardware and compare the results of the BRGC and binary codes with quantum Fourier transform. BRGC outperforms both the binary code  and QFT as it matches the exact result, within the 95\% confidence interval, while the others do not agree with the theoretical values. In future work we hope to study multi-particle systems in higher dimensions with gray code and also study fermionic and bosonic statistics.


\section{ACKNOWLEDGEMENTS}
We thank Alessandro Roggero for useful discussions and suggestions.

Lawrence Berkeley National Laboratory (LBNL) is operated by The Regents of
the University of California (UC) for the U.S. Department of Energy (DOE) under
Federal Prime Agreement DE-AC02-05CH11231.
This material is based upon work supported by the U.S. Department of Energy,
Office of Science, Office of Nuclear Physics,
Quantum Horizons: QIS Research and Innovation for Nuclear Science
under Field Work Proposal NQISCCAWL (CCC, KSM, YW, AWL).
ER acknowledges the NSF N3AS Physics Frontier Center, NSF Grant No. PHY-2020275, and the Heising-Simons Foundation (2017-228).
YW is supported under iTHEMS fellowship.

%

\appendix

\section{Circuit Compression}
\label{app:circuit_compression}
\begin{figure}[ht]
\begin{tabular}{c}
\includegraphics[scale=0.35]{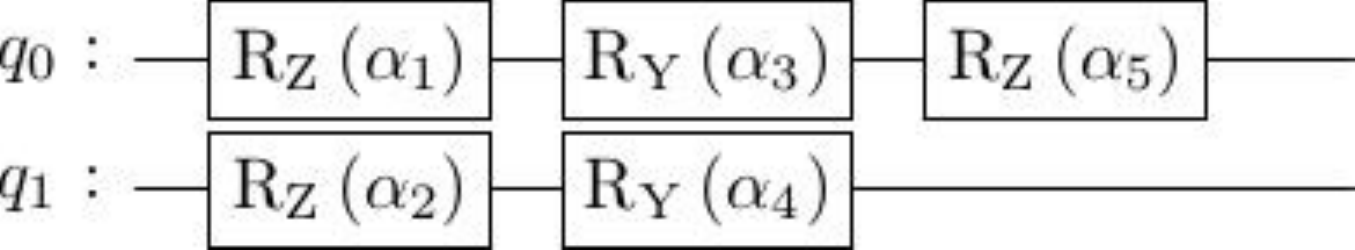}\\
(a) BRGC\\ \\
\includegraphics[scale=0.35]{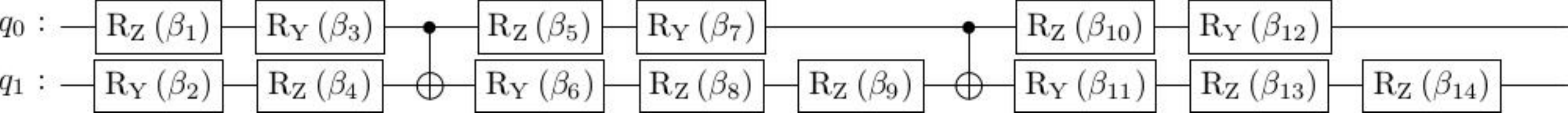}\\
(b) Binary\\ \\
\includegraphics[scale=0.35]{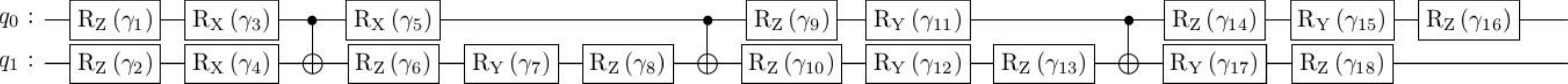}\\
(c) QFT
\end{tabular}
\caption{Circuit implementation for the two encodings and QFT for $n=2$. At first we decompose each circuit into the basis gates that can be compiled on the IonQ hardware, then compress it to an equivalent shorter circuit up to an overall phase.}
\label{fig:circuits_compressed}
\end{figure}

The respective circuits were compiled with the IBM
Qiskit library~\cite{qiskit,ibm}, and compressed with the transpiler provided. Single qubit gates were simplified with Euler decompositions, while for two and higher qubit gates we employed recent algorithms from the literature~\cite{Iten:2016,Cross:2019}. The resulting circuits are expressed as depicted in Fig.~\ref{fig:circuits_compressed}.
The form of the resulting circuits is independent of the number of Trotterization steps. As such, the addition of steps only serves to change the angles of the single qubit gates. While we do not expect these features to persists for higher number of qubits, these simplifications allow us to perform a complete adiabatic evolution schedule on NISQ hardware for small systems.

\onecolumngrid

\twocolumngrid

\end{document}